\documentclass[11pt]{article}
\usepackage[margin=1in]{geometry}
\usepackage{graphicx}
\usepackage[utf8]{inputenc}
\usepackage{lineno}

\usepackage[symbol]{footmisc}

\def\beq{\begin{equation}}
\def\eeq#1{\label{#1}\end{equation}}
\def\eeqn{\end{equation}}

\def\beqa{\begin{eqnarray}}
\def\eeqa#1{\label{#1}\end{eqnarray}}
\def\eeqan{\end{eqnarray}}

\let\bar=\overbar

\def\Dslash{\not{\hbox{\kern-4pt $D$}}}
\def\dslash{\not{\hbox{\kern-2pt $\del$}}}

\def\msb{{\bar{\ssstyle M \kern -1pt S}}}

\def\Title#1{\begin{center} {\Large {\bf #1} } \end{center}}
\def\Author#1{\begin{center} {\normalsize {\sc #1} } \end{center}}
\def\Institution#1{\begin{center} {\normalsize {\it #1} } \end{center}}
\def\Abstract#1{\noindent {\normalsize {\bf Abstract:} {\normalfont #1}}}
\def\Conference{\vspace{4mm}\begin{raggedright} {\normalsize {\it Talk presented at the 2019 Meeting of the Division of Particles and Fields of the American Physical Society (DPF2019), July 29--August 2, 2019, Northeastern University, Boston, C1907293.} } \end{raggedright}\vspace{4mm}}

\begin{document}

%
%

\Title{Anisotropy of Cosmic Ray Fluxes Measured with the Alpha Magnetic Spectrometer on the ISS}
\Author{M. Molero \footnote[1]{Speaker} \footnote[2]{on Behalf of the AMS Collaboration} }
\Institution{Centro de Investigaciones Energéticas, Medioambientales y Tecnológicas (CIEMAT)\\ Departamento de Investigación Básica, E-28040\\ Madrid, Spain\\ miguel.molero@ciemat.es}

\Author{J. Casaus, C. Mana, M.A. Velasco}

\Institution{Centro de Investigaciones Energéticas, Medioambientales y Tecnológicas (CIEMAT)\\ Departamento de Investigación Básica, E-28040\\ Madrid, Spain}

\Author{I. Gebauer, M. Graziani}

\Institution{Karlsruhe Institute of Technology, Institute for Experimental Particle Physics, D-76131\\ Kalsruhe, Germany}

\Author{M. Gervasi, G. La Vacca, P.G. Rancoita}

\Institution{INFN sez. Milano-Bicocca, Piazza della Scienza, 3 - 20126 Milano, Italy \\ Physics Department, University of Milano-Bicocca, Piazza della Scienza, 3 - 20126, Milano, Italy}

\Abstract{A measurement of the dipole anisotropy in galactic coordinates for different charged cosmic rays has been performed with the Alpha Magnetic Spectrometer (AMS) onboard the International Space Station (ISS). Results are presented for the first 7.5 years of data taking for protons, Helium, Carbon and Oxygen, and 6.5 years for positrons and electrons. All the species are found to be consistent with isotropy and upper limits to the dipole amplitude have been computed. In particular, for energies above 16 GeV a limit of $\delta < 1.9 \%$ and $\delta < 0.5 \%$ at the 95\% C.I. is found for positrons and electrons respectively. For rigidities above 200 GV a limit of $\delta < 0.38\%$, $\delta < 0.36\%$, $\delta < 1.9\%$ and $\delta < 1.7\%$ is obtained for protons, Helium, Carbon and Oxygen.}

\Conference

%
%

\section{Introduction}

\hspace{0.45cm} Precise measurements of the cosmic ray (CR) fluxes have been performed by AMS since its installation onboard the ISS in 2011. The results have revealed unpredicted features in their spectra that cannot be accounted within 
the current understanding of the production and propagation of galactic CRs.

\par
On the one hand, the proton \cite{ProtonFlux} and nuclei \cite{HeliumFlux,NucleiFlux} spectra deviate from a single power law and the spectral index progressively hardens above $\sim$ 200 GV. These effects may reveal the existence of local sources or a change in their propagation mechanisms.

\par
 On the other hand, the positron flux shows an excess above $\sim$ 25 GeV, followed by a sharp drop off above 284 GeV with a finite exponential energy cutoff at 810 GeV \cite{PositronFlux}. The observations cannot be explained by a pure secondary origin and for most of the explanations, the inclusion of primary sources is required; typically, being classified in two escenarios: dark matter and astrophysical sources \cite{AstroOriginPositron,DarkMattvsAstro}.
\par
 Furthermore, the electron flux shows an excess above $\sim$ 42 GeV which is well described in the entire energy range by using the sum of 2 power law components. Contrary to the positron flux, the electron flux does not have an energy cutoff below 1.9 TeV, and the nature of the electron excess indicates a different origin than from the positron one \cite{ElectronFlux}.

\par
In all cases, the contribution of nearby sources may induce some degree of anisotropy in the arrival directions of the differenct CR species. Therefore, its measurement would support or disfavor some of the aforementioned scenarios.

\section{The AMS-02 Detector}

\hspace{0.45cm} AMS-02 is a multipurpose particle physics detector installed onboard the ISS since the 19th of May 2011. It has been designed to carry out precise measurements of charged CRs in the GeV-TeV energy range. The detector has continuously collected data since its installation, with more than 140 billion of events recorded in more than 8 years. The end of the ISS is currently planned for 2024 and AMS will continue taking data until then.
\par
The detector consists of different sub-detectors that measure the charge, energy, momentum ($p$), or rigidity ($R=p/Z$) independently. The key elements used for the present analysis are the following: a silicon tracker (STD) with an inner tracker (L2-L8) inside a permanent magnet and two outer layers (L1 and L9), one at the top and the other at the bottom of the detector; a transition radiation detector (TRD); a time of flight (TOF); a ring imaging Cherenkov detector (RICH); and an electromagnetic calorimeter (ECAL). A detailed description can be found in \cite{Detector}.


\section{Selection}

\hspace{0.45cm} For all the different species used in this analysis a cut based selection has been applied.
\par
In the case of the positron and electron samples, the selected events are required to be relativistic downward-going particles with measured velocity $\beta \sim 1$, to have a reconstructed shower in the ECAL, with a matched track in the tracker and the TRD, charge consistent with $Z=1$, and quality criteria are applied to ensure good accuracy of the track reconstruction. Further cuts on the TRD and ECAL estimators as well as good energy-momentum matching are used to reduce the proton background below the percent level ~\cite{Detector,PosFrac1}. The total samples include $9.9\times10^4$ positrons and $1.3\times10^6$ electrons above 16 GeV. The selected events are grouped into 5 cumulative energy ranges from 16 to 350 GeV, with minimum energies 16, 25, 40, 65 and 100 GeV, respectively.

\par
For the proton, Helium, Carbon and Oxygen selection, events are required to be downward-going particles and to have reconstructed track in the inner passing through the L1 (L1-L8) with an additional hit in the L9 (L1-L9) for protons. Quality criteria cuts in the reconstruction of the track are applied. Finally, consistency with the charge of the respective species is required \cite{ProtonFlux,HeliumFlux,NucleiFlux}. The total number of selected events is $1.3\times10^8$, $1.0\times10^8$, $2.9\times10^6$ and  $2.8\times10^6$ for protons, Helium, Carbon and Oxygen and for $R > 18$ GV, where in this case the protons and nuclei are grouped into 9 cumulative ranges of minimum rigidity: 18, 30, 45, 80, 150, 200, 300, 500, 1000 GV.

\section{Methodology}

\hspace{0.45cm} The measurement of a large scale anisotropy consists in the determination of the directionality of the CR flux in a coordinate system. For this analysis, all the results will be presented in galactic coordinates.

\par
 Directional fluctuations can be computed by comparing the map of selected events with a reference map; any deviation from this map might be regarded as a signal. The reference map describes the response of the detector to an isotropic flux and, therefore, a very precise understanding of the detector's behavior is required. In particular, it is necessary to understand the geographical dependences of the efficiencies and its projection in galactic coordinates; if not taken into account properly some spurious signals might be detected.

\par
To compute large scale anisotropies the flux is expanded in a basis of spherical harmonics,

\begin{equation}
f(\theta, \phi) = \sum_{l=0}\sum_{m=-l}^{m=+l}a_{lm}Y_{lm}(\theta,\phi)
\end{equation}
\\
where the $Y_{lm}$ are the real spherical harmonics of degree $l$ and order $m$, with $l=0,1,2$.. and $m=0$,$\pm1$,$\pm2$,...,$\pm$ $l$ and $a_{lm}$ are the coefficients of the expansion, which determine the degree of the anisotropy.
\par
The large scale anisotropy is described at first order by a dipole ($l=1$) and its projection onto 3 orthogonal directions (East-West, North-South and Forward-Backward). For the case of galactic coordinates the North-South (NS) direction is perpendicular to the galactic plane, the Forward-Backward (FB) is pointing to the galactic center, and finally the East-West (EW) completes the righ-handed coordinate system and is contained into the galactic plane. The three dipole components can be defined as
\begin{equation}
  \rho_{EW}=\sqrt{\frac{3}{4\pi}}a_{1-1}\hspace{0.45cm}    ; \hspace{0.45cm}   \rho_{NS}=\sqrt{\frac{3}{4\pi}}a_{10} \hspace{0.45cm}      ;    \hspace{0.45cm}  \rho_{FB}=\sqrt{\frac{3}{4\pi}}a_{11}
\end{equation}
Finally, the dipole amplitude can be computed as follows
\begin{equation}
\delta=\sqrt{\rho^2_{EW}+\rho^2_{NS}+\rho^2_{FB}}
\end{equation}

\section{Anisotropy of protons and light nuclei}
\begin{figure}[!h]
\centering
\includegraphics[height=2.2in, width=3.2in]{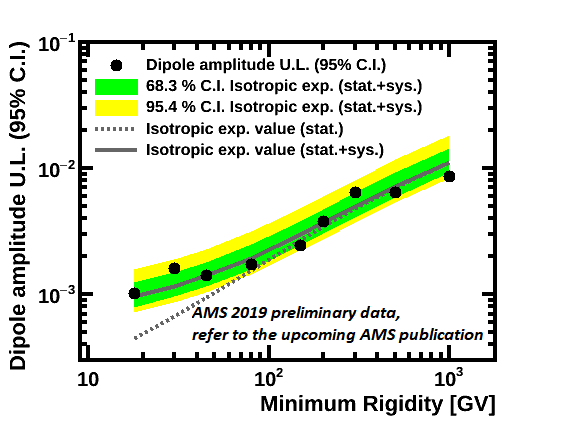}
\hspace{-0.2in}
\includegraphics[height=2.2in]{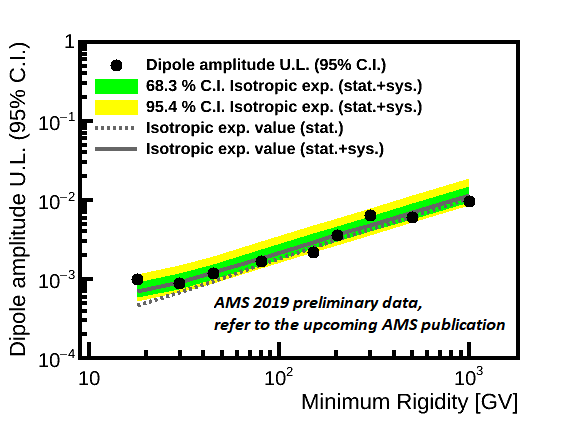}
\caption{Upper limits on the dipole amplitudes for protons (\emph{left}) and Helium (\emph{right}) . The points show the AMS measurement, the solid (\emph{dashed}) lines correspond to the isotropic expectation with $(without)$ systematic uncertainties, and the bands correspond to the 68.3\% and 95.4\% regions. }
\label{fig:protonsUL}
\end{figure}

\begin{figure}[!h]
\centering
\includegraphics[height=2.2in]{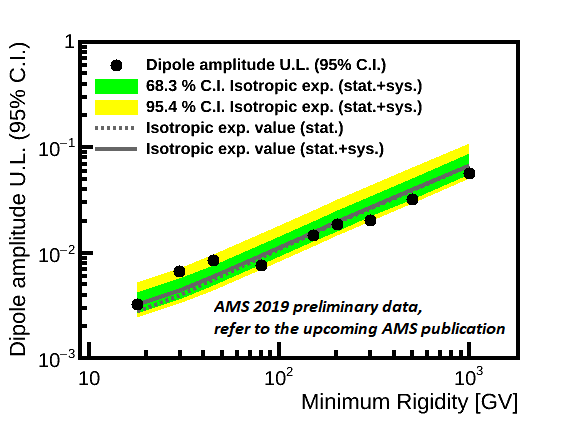}
\includegraphics[height=2.2in]{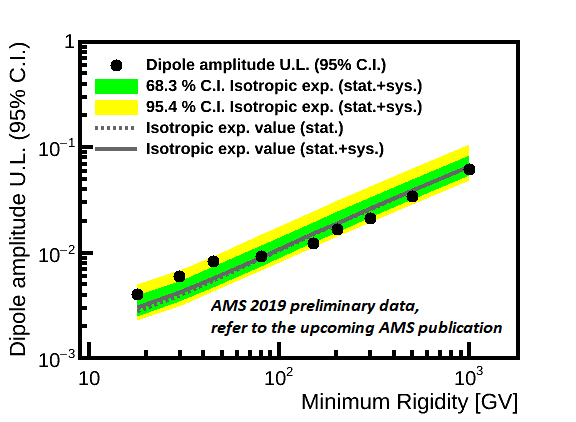}
\caption{Upper limits on the dipole amplitudes for Carbon (\emph{left}) and Oxygen (\emph{right}).}
\label{fig:nucleiUL}
\end{figure}

\hspace{0.45cm} The method was applied to compute the anisotropy of protons, Helium, Carbon and Oxygen and no deviations from isotropy were found. Consequently, upper limits at the 95\% C.I. have been set for all the different species and rigidity ranges of the analysis, Fig. \ref{fig:protonsUL} and Fig. \ref{fig:nucleiUL}. The limits obtained above 200 GV are $\delta < 0.38\%, \delta < 0.36\%, \delta < 1.9\%$ and $\delta < 1.7\%$ for protons, Helium, Carbon, and Oxygen respectively. In particular, in the case of protons for $R> 70$ GV the measurement is limited by statistics. At lower rigidities, the systematics on the efficiency corrections limit the sensitivity to the per mil level (0.1\%).

\pagebreak

\section{Anisotropy of positrons and electrons}
\begin{figure}[h]
\centering
\hspace{-0.25in}
\includegraphics[height=2.05in]{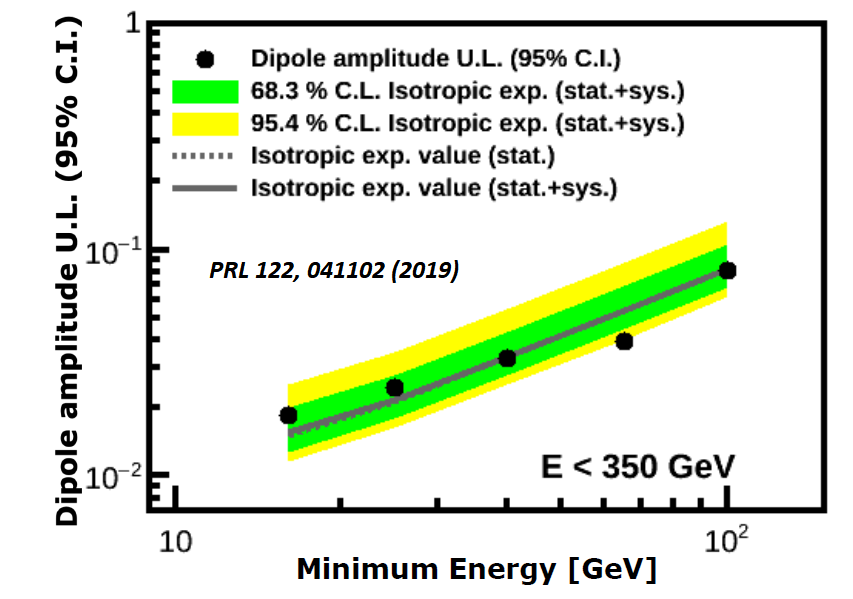}
\hspace{0.1in}
\includegraphics[height=2.05in]{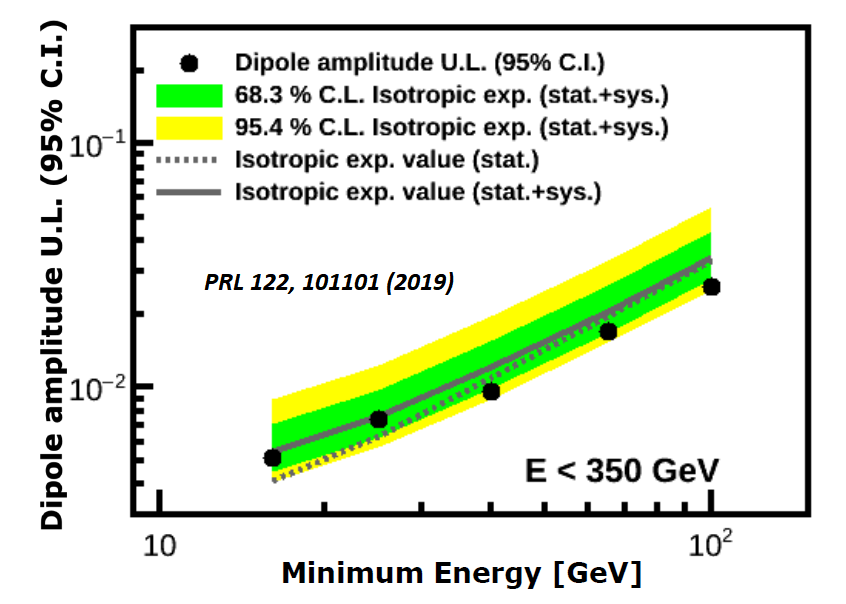}
\caption{Upper limits on the dipole amplitudes for positrons (\emph{left}) and electrons (\emph{right}).}
\label{fig:PosEleUL}
\end{figure}

\hspace{0.45cm} Following the same procedure for positrons and electrons the dipole components are also consistent with isotropy. Limits to the dipole amplitude at the 95\% C.I. can be set for the different energy ranges, Fig. \ref{fig:PosEleUL}. In particular, for energies above 16 GeV a limit of $\delta < 1.9 \%$ and $\delta < 0.5 \%$ is obtained for positrons and electrons.

\section{Conclusions}
\hspace{0.45cm} Measurement of the dipole anisotropy in galactic coordinates for protons, Carbon, Oxygen, positrons and electrons have been performed by the AMS-02 detector. The results are presented for the first 7.5 years of data taking for protons and light nuclei, and 6.5 years for the positrons and electrons. No deviations with respect to isotropy have been found, which allow to set limits at the 95\% C.I. to the dipole amplitude for the different species presented in this analysis. In particular, for $R > 200$ GV limits of $\delta < 0.38\%, \delta < 0.36\%, \delta < 1.9\%$ and $\delta < 1.7\%$ are obtained for protons, Helium, Carbon and Oxygen. For positrons and electrons, limits of $\delta < 1.9 \%$ and $\delta < 0.5 \%$ are found.

\end{document}